\documentclass[conference]{IEEEtran}
\IEEEoverridecommandlockouts
\usepackage{bm}
\usepackage{cite}
\usepackage{amsmath,amssymb,amsfonts}
\usepackage{algorithmic}
\usepackage{graphicx}
\usepackage{textcomp}
\usepackage{xcolor}
\usepackage{multirow}
\usepackage{tikz}
\usepackage[T1]{fontenc}
\def\BibTeX{{\rm B\kern-.05em{\sc i\kern-.025em b}\kern-.08em
    T\kern-.1667em\lower.7ex\hbox{E}\kern-.125emX}}
    
\newcommand\copyrighttext{%
  \footnotesize \textcopyright 2020 IEEE. Personal use of this material is permitted.
  Permission from IEEE must be obtained for all other uses, in any current or future
  media, including reprinting/republishing this material for advertising or promotional
  purposes, creating new collective works, for resale or redistribution to servers or
  lists, or reuse of any copyrighted component of this work in other works.}
\newcommand\copyrightnotice{%
\begin{tikzpicture}[remember picture,overlay]
\node[anchor=south,yshift=10pt] at (current page.south) {\fbox{\parbox{\dimexpr\textwidth-\fboxsep-\fboxrule\relax}{\copyrighttext}}};
\end{tikzpicture}%
}    
    
\begin{document}

\title{PGD-UNet: A Position-Guided Deformable Network for Simultaneous Segmentation of Organs and Tumors\\
}
\author{\IEEEauthorblockN{Ziqiang Li}
\IEEEauthorblockA{\textit{School of Automation}\\
\textit{Southeast University}\\
No.2 Sipailou, Nanjing, China, 210096\\
zql@seu.edu.cn}
\and
\IEEEauthorblockN{Hong Pan}
\IEEEauthorblockA{\textit{Department of Computer Science and Software Engineering}\\
\textit{Swinburne University of Technology}\\
John St, Hawthorn, VIC 3122, Australia\\
hpan@swin.edu.au}
\and
\IEEEauthorblockN{Yaping Zhu}
\IEEEauthorblockA{\textit{School of Information and Communication Engineering} \\
\textit{Communication University of China}\\
No.1 Dingfuzhuang East Street, Beijing, China, 100024\\
zhuyaping@cuc.edu.cn}
\and
\IEEEauthorblockN{A. K. Qin}
\IEEEauthorblockA{\textit{Department of Computer Science and Software Engineering}\\
\textit{Swinburne University of Technology}\\
John St, Hawthorn, VIC 3122, Australia\\
kqin@swin.edu.au}
}

\maketitle
\copyrightnotice

\begin{abstract}
Precise segmentation of organs and tumors plays a crucial role in clinical applications. It is a challenging task due to the irregular shapes and various sizes of organs and tumors as well as the significant class imbalance between the anatomy of interest (AOI) and the background region. In addition, in most situation tumors and normal organs often overlap in medical images, but current approaches fail to delineate both tumors and organs accurately. To tackle such challenges, we propose a position-guided deformable UNet, namely PGD-UNet, which exploits the spatial deformation capabilities of deformable convolution to deal with the geometric transformation of both organs and tumors. Position information is explicitly encoded into the network to enhance the capabilities of deformation. Meanwhile, we introduce a new pooling module to preserve position information lost in conventional max-pooling operation. Besides, due to unclear boundaries between different structures as well as the subjectivity of annotations, labels are not necessarily accurate for medical image segmentation tasks. It may cause the overfitting of the trained network due to label noise. To address this issue, we formulate a novel loss function to suppress the influence of potential label noise on the training process. Our method was evaluated on two challenging segmentation tasks and achieved very promising segmentation accuracy in both tasks.
\end{abstract}


\section{Introduction}
Medical imaging, e.g., magnetic resonance imaging (MRI) and computed tomography (CT), plays a crucial role in cancer diagnosis and treatment decision, where precise and robust segmentation of organs and tumors in medical images is of great value. Benefitting from its powerful feature representation capability, deep learning has achieved breakthrough performance in many medical image analysis tasks such as pulmonary nodule detection~\cite{liao2019evaluate} and brain tumor segmentation~\cite{myronenko20183d}. With the advent of convolutional neural networks (CNNs), abundant work on medical image segmentation has been proposed, including skip-connections~\cite{ronneberger2015u}, distance transform maps~\cite{kervadec2018boundary}, attention mechanisms~\cite{oktay2018attention}, etc. The performance on some simple tasks has reached the level of radiologists. However, there remains many challenges to overcome in order to meet the practical requirements in the segmentation of organs and tumors. Specifically, tumor tissues tend to have irregular shapes due to their invasive nature, leading to shape variations. In most cases, tumors often overlap with organs, which causes obstacle for accurate segmentation of organs and tumors simultaneously. There may exist large size variations between inter- and intra- subjects caused by different cancer stages and inherent inter-category differences. Radiologist's subjective annotations and the uncertainty of malignant tumor boundaries may introduce label noise. Extreme class imbalance between the AOI and the background region also cause difficulty for medical image segmentation.

To tackle the aforementioned challenges, some innovative building blocks have been incorporated into conventional CNNs to improve its robustness to shape variations. Dai et al.~\cite{dai2017deformable} firstly introduced deformable convolution. By adding additional offsets to the regular grid sampling locations of convolution kernels, it enhances CNN's capability of modeling geometric transformation. Despite the improved modeling of geometric transformation, there remain some issues in deformable convolution. First of all, deformable convolution requires precise position information to calculate the offset, which is conflicted with CNN's position insensitivity (a.k.a. translation invariance). On the other hand, the offsets are learned from the preceding feature map, although it is hard to guarantee that appropriate offsets are learned with the same receptive field. In this work, we propose a position-guided deformable network, namely PGD-UNet, to deal with the deformation of anatomical structures, such as organs and tumors. It consists of a U-Net backbone incorporated with deformable convolution and an auxiliary localization path. The localization path explicitly introduces position information to guide deformable convolution, which effectively improves the capability of modeling geometric transformation. Meanwhile, in order to accommodate the structures of various sizes in an image, we use Atrous Spatial Pyramid Pooling (ASPP)~\cite{chen2017rethinking} as the bottleneck layer to extract multi-scale features.

In medical image segmentation, small structures also cause class imbalance, where the anatomy of interest only occupies a very small portion of the image. For example, in the bladder MRI image used in our experiments, the tumor region is composed of only 0.63\% of all pixels. Existing approaches to addressing class imbalance can be categorized into two groups, i.e., multi-stage cascaded CNNs and re-weighting the losses contributed from different classes. The former approach detects the AOI and then segments out the target from that particular region. This approach is computational excessive and not easy to be extended to multi-class segmentation. The focal loss~\cite{lin2017focal} was proposed to make the network to focus on hard-to-classify samples which influence more on classification performance. However, mislabeled samples and hard-to-classify samples are prone to be confused. In this work, we propose a novel noise suppression focal loss to suppress the effect of mislabeled samples and thus prevent the network from overfitting.

We test the proposed approach on two challenging medical segmentation tasks: bladder tumors segmentation in MRI and pancreas tumors segmentation in CT. Both the bladder dataset and the pancreas dataset from the Medical Segmentation Decathlon Challenge (MSD)~\cite{simpson2019large} need segment organs and tumors simultaneously, and suffer from class imbalance due to large (background), medium (pancreas, bladder wall) and small (tumor) structures. Experimental results show that our approach can improve on prediction accuracy on both datasets and achieve state-of-the-art performance.

\section{Related Work}

\subsection{Spatial Transformation}
Effective modeling of spatial transformation is a key challenge in visual recognition. The typical method is to augment the training samples with sufficient desired variations through translation, rotation, scaling, etc., which is simple but laborious. Furthermore, some transformation-invariant features are designed, such as scale-invariant feature transform (SIFT)~\cite{lowe2004distinctive} and local binary patterns (LBP)~\cite{ojala1996comparative}. Nevertheless, such handcrafted features need expert knowledge for careful design, but lack sufficient generalization power to different domains. Although deep CNNs have powerful representation capabilities, its invariance still implicitly relies on data augmentation, parameter sharing, and pooling operations etc. Spatial transformer networks (STN)~\cite{jaderberg2015spatial} is the first work that model geometric transformations in a computational and parametric manner. The spatial transformer module dynamically learns a set of global affine transformation parameters from feature map, and then transmits the transformed feature map to subsequent layers to simplify recognition. Instead of performing global affine transformations, deformable convolution~\cite{dai2017deformable} learns a dense kernel-wise offset, which endows ordinary convolution operations the flexibility to adapt to objects with more complex geometric transformations. Our work addresses two drawbacks of deformable convolution: position insensitivity and local receptive field.

\subsection{Class Imbalance}
Class imbalance is quite common in medical image segmentation. A general solution is to exploit multi-stage cascaded CNNs~\cite{roth2018spatial}, which directly eliminates most of the background through the first detection stage among the pipeline. Another genre is the re-weighting method. Cross-Entropy (CE) based weight loss~\cite{long2015fully,kamnitsas2017efficient,ronneberger2015u} re-weights the different classes according to the frequency of corresponding labels. Focal loss~\cite{lin2017focal} further integrates the difficulty of the sample for weighting. Gradient harmonizing mechanism (GHM) loss~\cite{li2019gradient} directly calculates the gradient distribution of each batch, and alleviates class imbalance by flattening the gradient. Dice loss~\cite{sudre2017generalised} based on regional integration is commonly used to handle unbalanced medical segmentation. Kervadec et al.~\cite{kervadec2018boundary} proposed a boundary loss, which formulates a distance metric on the space of contours to mitigate the difficulties of regional losses.

\subsection{Label Noise}
In medical image analysis, the presence of label noise is quite common due to the uneven image quality and the high clinical expertise required for annotation. To solve this problem, Minimal annotation training~\cite{matuszewski2018minimal} is developed to segment microscopy virus particles with coarse annotations. This method first generates masks for suspected noise regions, then ignores these regions when calculating dice similarity loss. In reference~\cite{dgani2018training}, a noise layer is added to the end of CNNs for breast lesion detection. Noise layer can be considered as a transformation matrix of noise and true labels, which are optimized with a combination of expectation maximization (EM) and error back-propagation. Some methods are based on sample re-weighting and feature consistency.

\begin{figure*}[]
\centering
\includegraphics[scale=0.6]{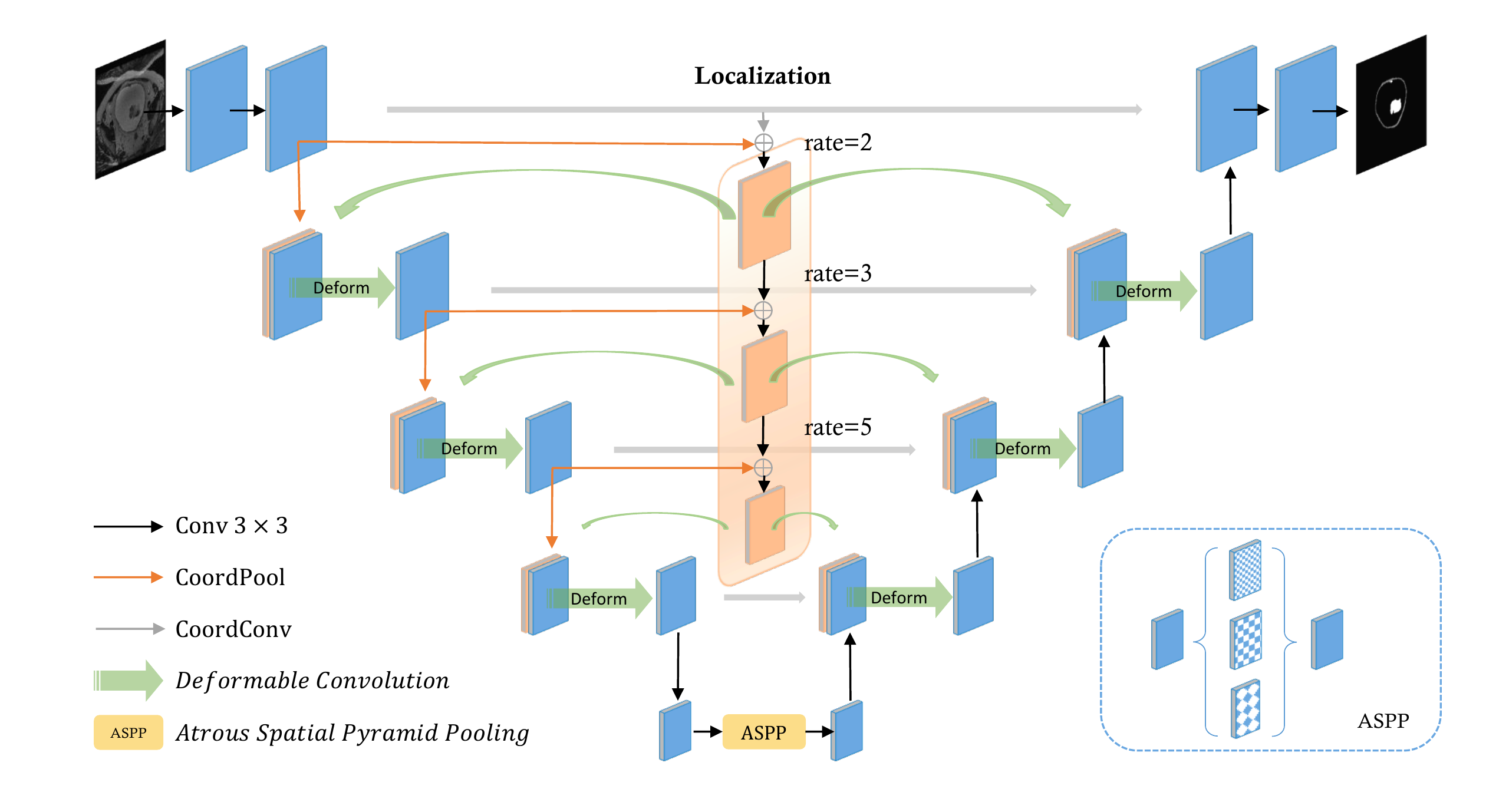}
\caption{The network architecture of our proposed PGD-UNet for medical image segmentation. Blue and orange blocks represent feature maps of the backbone and localization path, respectively}
\label{arch}
\end{figure*}

\section{Method}
\subsection{Network Architecture}
Fig.~\ref{arch} illustrates the architecture of our PGD-UNet, where U-Net is adopted as the backbone. The backbone consists of an encoding path to extract semantic information and a symmetric decoding path for recovery. To accommodate irregular and complex geometric variations of organs and tumors, deformable convolutions are embedded into the middle three blocks of the two paths. Nevertheless, the deformable convolution operator (DCO) requires accurate position information to generate coordinate offset and mask, which is agnostic in the plain convolution feature map due to CNN's inherent translation invariance. Consequently, we introduce an auxiliary position-sensitive localization path to provide DCO with additional position information. The localization path does not share the parameters of the encoding path, and position information is added by the form of coordinates. To handle size variations between organs and tumors, as well as the tumors of different stages, we adopt Atrous Spatial Pyramid Pooling (ASPP) as a bottleneck layer so that the network can represent multiple structures of different sizes simultaneously by extracting features with different receptive fields.

\subsection{Position-Guided Deformable Convolutional Layers}
An essential strength of our proposed segmentation network is to model spatial transformations. To achieve this, the deformable convolution is introduced to enable a dense pixel-wise deformation. In addition, a novel position-aware path is included to further improve the current deformation paradigm.

\subsubsection{Deformable Convolution}
The standard convolution can be regarded as using a regular grid $\mathcal{R}$ to sample over the input $x$, and then sum the sampled values weighted by $w$. For example, a $3 \times 3$ kernel is defined as:
$$\mathcal{R}=\{(-1,-1),(-1,0), \ldots,(0,1),(1,1)\}$$

The value at location $P_{0}$ on the output feature map $y$ is calculated as:
\begin{equation}
y\left(P_{0}\right)=\sum_{P_{n} \in R} w\left(P_{n}\right) \cdot x\left(P_{0}+P_{n}\right)
\label{eq1}
\end{equation}
where $w$ is the kernel weight and $P_{n}$ enumerates the sampling location of $\mathcal{R}$.

The deformable convolution adjusts the position of grid sampling cell with offset $\Delta P_{n}$ and multiplies each offset sampling cell by a modulated weight $\Delta m_{n}$, where $n = 1,2,...,N$, and $N$ is equal to the number of cells in the grid $\mathcal{R}$. For deformable convolution, Eq.~\ref{eq1} becomes
\begin{equation}
y\left(P_{0}\right)=\sum_{P_{n} \in R} w\left(P_{n}\right) \cdot x\left(P_{0}+P_{n}+\Delta P_{n}\right) \cdot \Delta m_{n}
\end{equation}
The offset $\Delta P_{n}$ is a pair of learnable parameters with unconstrained range, while mask $\Delta m_{n}$ varies in $[0,1]$. The $x\left(P_{0}+P_{n}+\Delta P_{n}\right)$ is computed via bilinear interpolation.

\begin{figure}[]
\centering
\includegraphics[scale=0.5]{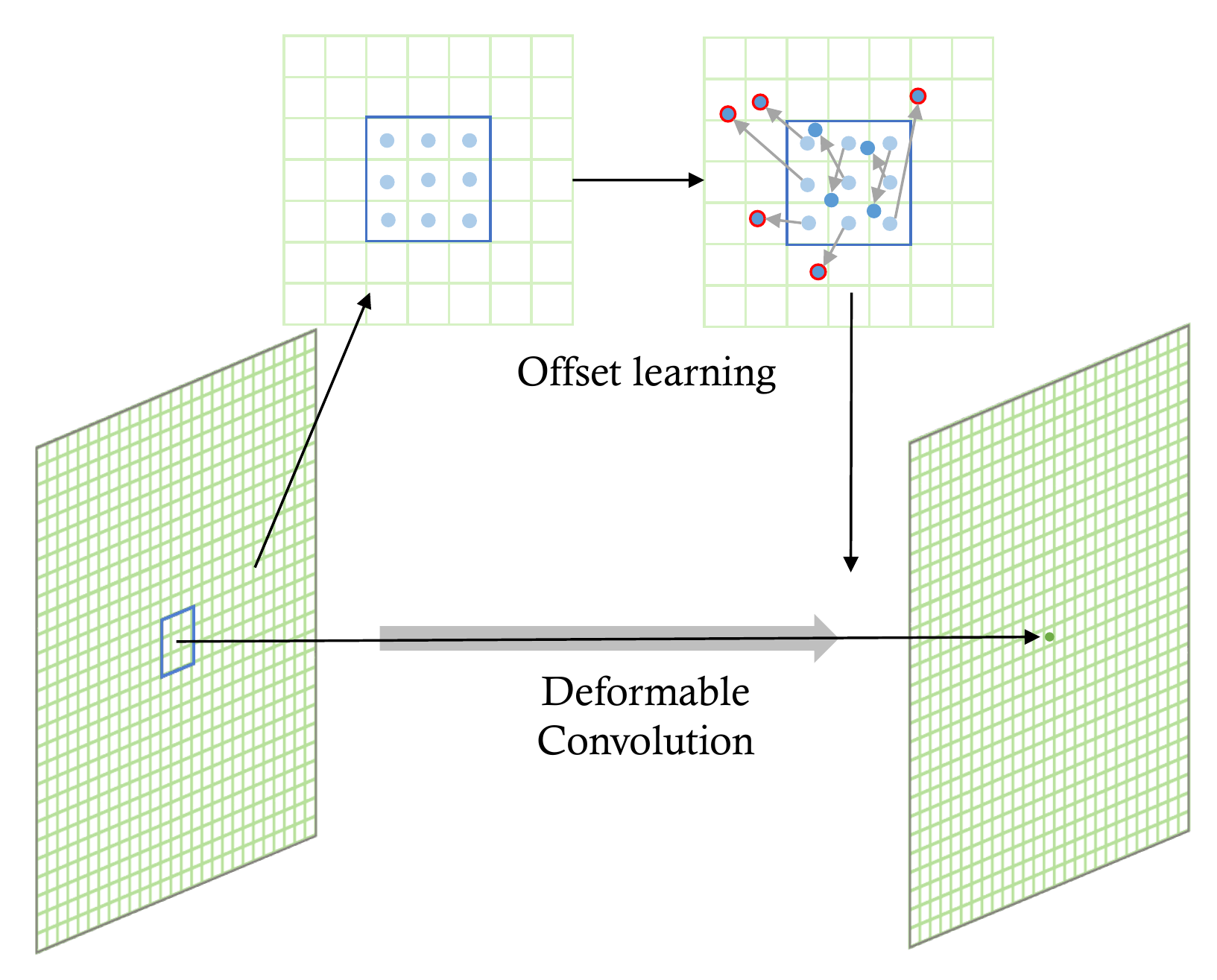}
\caption{Deformable convolution with $3\times3$ kernel.}
\label{deform}
\end{figure}

As illustrated in Fig.~\ref{deform}, both offset and mask are learned through an additional convolution layer with the same input feature map $x$, which has the same kernel size and dilation as the deformable convolution in the main branch. For example, a $3 \times 3$ deformable kernel with dilation 1 samples over the input feature map with a $3 \times 3$ shifted grid $\mathcal{R}^\prime$, while the offsets are learned through a regular grid $\mathcal{R}$, shown in Fig.~\ref{deform}. Consequently, a natural problem is that when the shifted sampling point is outside the $3 \times 3$ regular grid (points with red outline in Fig.~\ref{deform}), it is agnostic that whether an appropriate offset can be learned, because the receptive field of this point has exceeded those calculate it (the normal spatial range of a 3x3 grid).

\subsubsection{Localization Path}
CNNs are generally considered to be position insensitive or translation invariance because features are extracted in a local manner. Nevertheless, recent studies exploring the interpretability of neural networks have shown that CNNs learn to encode position information within the feature maps implicitly, i.e., the neurons in deep layers know not only what they are representing, but also where they are. The success of position-dependent tasks (e.g. object detection and segmentation) also confirms this viewpoint. To evaluate the capability to encode position information of CNNs, Liu et al. \cite{liu2018intriguing} designed a simple coordinate mapping experiment. The results show that CNNs cannot recover the coordinates accurately. Therefore, CNNs can only learn a coarse position representation, but it is defective to calculate the accurate offset for deformable convolution. In this regard, we proposed an auxiliary localization path providing explicit position information to guide the offset computation and decouple semantic and position extraction.

\paragraph{Larger Receptive Field} As illustrated in Fig.~\ref{arch}, we stack three dilated convolution layers as the backbone of the localization path. To avoid the `gridding effect'~\cite{wang2018understanding}, we adopt $dilation rates = (2,3,5)$ for the three dilated convolution layers, respectively. The localization path takes the output feature map of the first block of UNet as input, which is the same as the subsequent layers in the encoder path. In order to maintain the same spatial resolution as the feature map at each block of the main branch, we adopt convolutions with $stride = 2$ for downsampling. Then the feature maps calculated by localization path are concatenated into the main branch along the channel dimension to guide the offset and mask calculation. As the stacked dilated convolutions employed in localization path introduce a larger receptive field than standard convolutions in encoding path, it helps avoid the above-mentioned problem of agnostic in shifted sampling point.

\paragraph{Position Sensitivity} To obtain appropriate offset, the localization path needs to be position sensitive. Consequently, we utilize the `CoordConv' operator~\cite{liu2018intriguing} to explicitly send the coordinates of each pixel in the image as additional information to the network. Specifically, before sending the feature map of the first block to the localization path, we add an `addCoord' layer. The `addCoord' layer generates the coordinates at $X$ and $Y$ axes for each pixel, and normalizes them to $[-1,1]$. The normalized coordinates are concatenated into the input feature map along the channel dimension. So the number of output channels will plus two.

\begin{figure}[]
\centering
\includegraphics[scale=0.5]{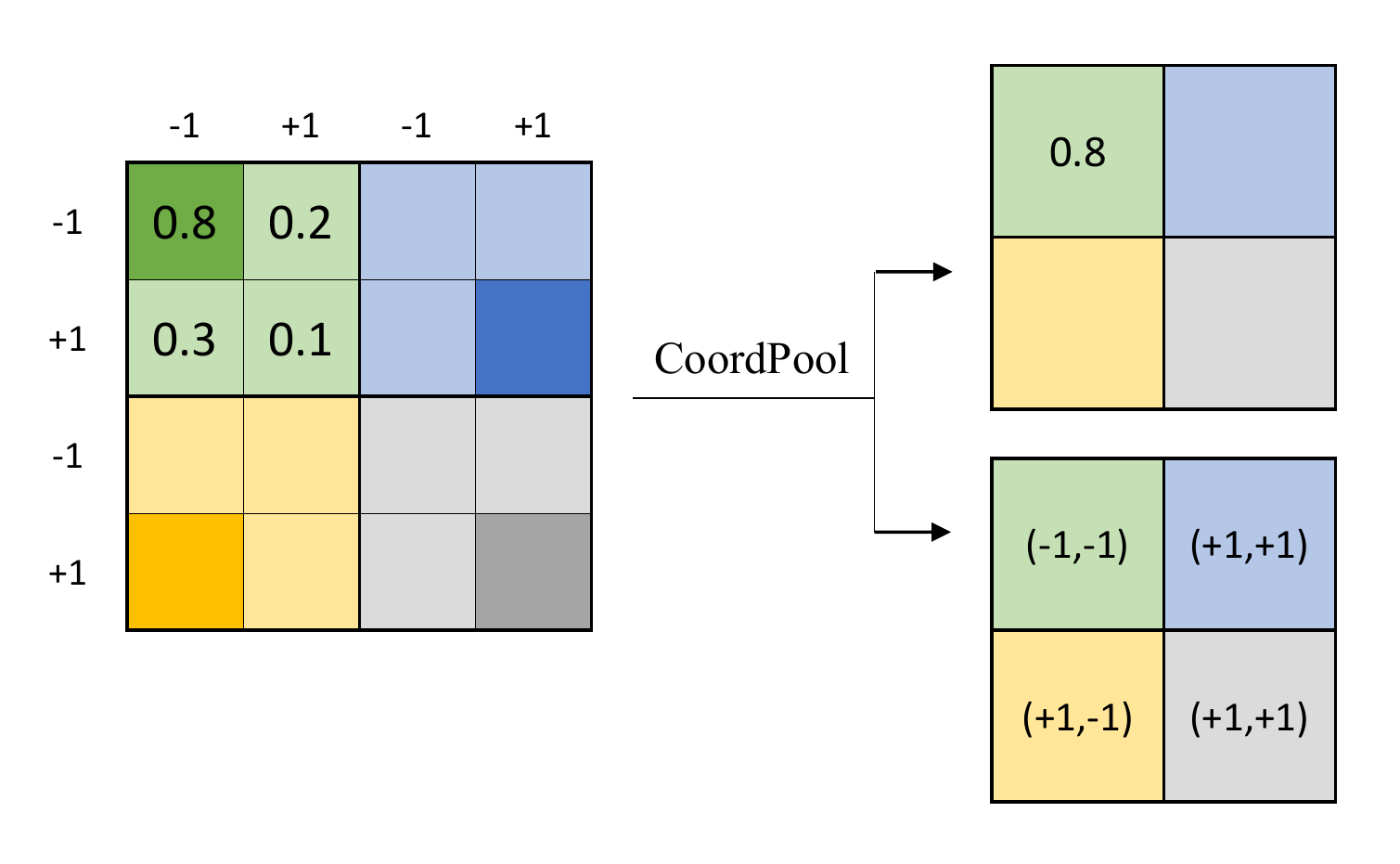}
\caption{CoordPool with $2\times2$ kernel, $2\times2$ strides. Each color represents a pool region}
\label{coordPool}
\end{figure}

Inspired by the work of Unpooling~\cite{zeiler2014visualizing}, we further propose a novel maximum pooling operation, called, CoordPool, to
perform normal \textit{max-pooling} operation while outputting the locations of the maxima within each pooling region. As illustrated in Fig.~\ref{coordPool}, the locations represent the coordinates of maxima in the pooling region, along $X$ and $Y$ axes. In our network, the locations of each block, output from CoordPool, is concatenated to the corresponding feature map in the localization path.

As we explicitly introduce the coordinate information into the network, hence PGD-UNet constructs a position-sensitive deformable convolution. In PGD-UNet, CoordPool preserves the spatial information lost by \textit{max-pooling} and passes it to the decoding path via skip-connections. In this way, our network has the capability of Unpooling.

\begin{figure*}[]
\centering
\includegraphics[scale=0.45]{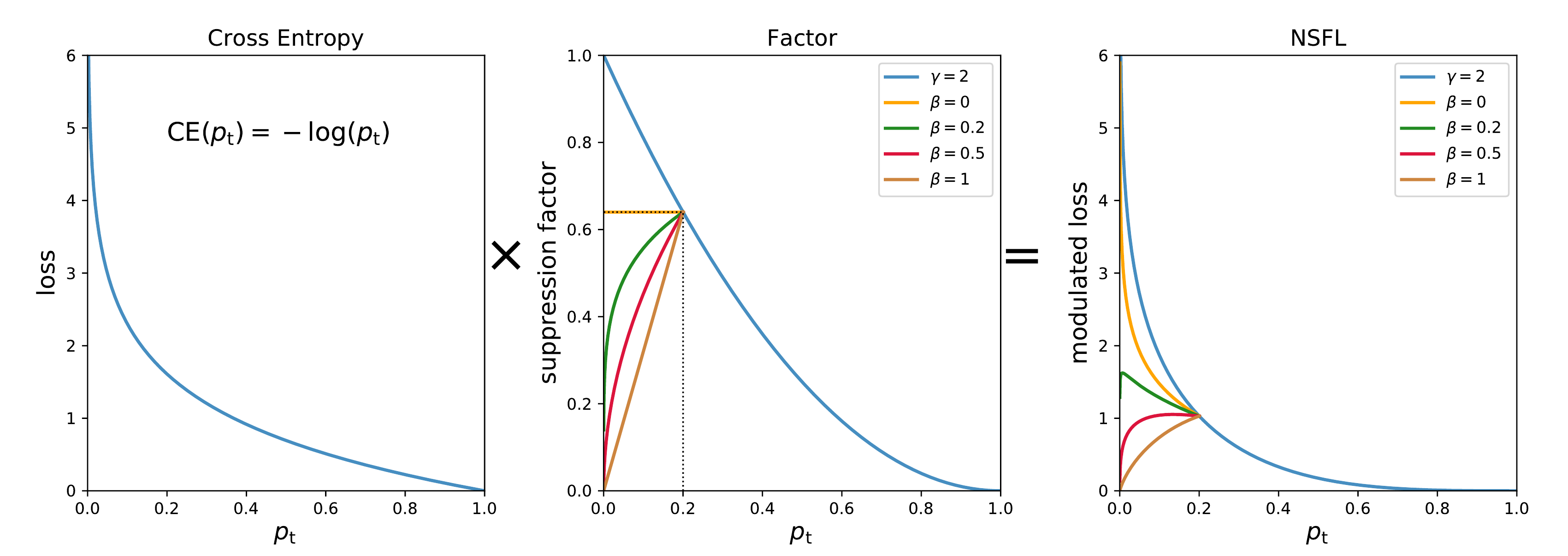}
\caption{Noise suppression focal loss. From left to right are the cross-entropy loss function, the modulating factor, and the final loss function, respectively.}
\label{loss}
\end{figure*}

\subsection{Noise Suppression Focal Loss}
Tumor segmentation is a difficult problem due to the following challenges: 1). malignant tumors usually have unclear boundaries; 2). the quality of images generated by different devices vary significantly; 3). manual delineation of tumors subject to inter- and intra-observer variations. All kinds of problems make label noise almost inevitable in medical images, which seriously affects the training process of neural networks. Firstly, during the initial phase of network convergence, neural networks tend to learn common features shared among the data samples~\cite{arpit2017closer}. At this point, the noise label will have a large error and appear as an outlier. Traditional loss functions, e.g., cross-entropy loss, will strengthen the penalty for noise, which causes the gradient to be dominated by mislabeled samples. Secondly, the proportion of tumor pixels in medical image is very small, which makes networks easily overfit the noise labels.

To solve this problem, we design a noise suppression focal loss to suppress the contribution of outliers to the gradient. In multi-class segmentation, the ground-truth of each pixel is encoded by a one-hot vector, where label $1$ represents the true class. Let $p_{t}$ denotes the predicted probability of the ground-truth class. The cross entropy (CE) loss can be written as:
\begin{equation}
\mathrm{CE}\left(p_{\mathrm{t}}\right)=-\log \left(p_{\mathrm{t}}\right)
\end{equation}

As shown in Fig.~\ref{loss}, difficult examples ($p_{t} \leqslant 0.5$) have greater losses than easy examples in CE loss. However, the difference of this magnitude can be overwhelmed easily in case of large class imbalance. Focal loss (FL)~\cite{lin2017focal} further amplifies this difference by adding a modulating factor $\left(1-p_{\mathrm{t}}\right)^{\gamma}$ to CE loss.
\begin{equation}
\mathrm{FL}\left(p_{\mathrm{t}}\right)=-\left(1-p_{\mathrm{t}}\right)^{\gamma} \log \left(p_{\mathrm{t}}\right)
\end{equation}

As our experiments will show, focal loss is very useful for dealing with extreme class imbalance. But at the same time, mislabeled samples also lie in low predicted $p_{t}$ regions and get large gradient. To alleviate the effects of noise, we design a piecewise focal loss, namely noise suppression focal loss (NSFL). Let $\epsilon$ denotes the piecewise parameter, NSFL replaces the modulating factor in focal loss with $\left(p_{\mathrm{t}}\right)^{\beta}$ when $p_{t}<\epsilon$.
\begin{equation}
\mathrm{NSFL}\left(p_{\mathrm{t}}\right)=\left\{\begin{array}{ll}
{-\frac{{\left(1-\epsilon\right)}^\gamma}{\epsilon^\beta} \left(p_{\mathrm{t}}\right)^{\beta} \log \left(p_{\mathrm{t}}\right),} & {p_{t}<\epsilon} \vspace{1ex} \\
{-\left(1-p_{\mathrm{t}}\right)^{\gamma} \log \left(p_{\mathrm{t}}\right),} & {p_{t} \geqslant \epsilon}
\end{array}\right.
\end{equation}

The $\beta$ varies in $[0,1]$, hence the replaced factor $\left(p_{\mathrm{t}}\right)^{\beta}$ suppresses gradient when $p_{t}$ is less than the threshold $\epsilon$. The degree of suppression depends on the value of $p_{t}$. When $\beta = 0$, it is equivalent to the factor being truncated, and when $\beta = 1$, the factor $\left(p_{\mathrm{t}}\right)^{\beta}$ becomes linear function, as shown in Fig.~\ref{loss}.

Furthermore, if the networks train from scratch, it is recommended to apply noise suppression focal loss after a few epochs because the prediction probability obtained by a randomly initialized network is meaningless. In our experiments, the average value of $p_{t}$ is used to decide when to switch to the noise suppression focal loss.

Finally, the overall loss function we formulate is a combination of weighted noise suppression focal loss and dice loss.
\begin{equation}
\mathcal{L}_{\text {all}}=\lambda \mathcal{L}_{\textit{NSFL}}+ (1-\lambda) \mathcal{L}_{\text {Dice}}
\end{equation}
where $\lambda$ is used to adjust the weight flexibly between two loss terms, according to the dataset.

\section{Experiments}

\subsection{Datasets}
To justify the effectiveness of our approach, two challenging tasks are evaluated, both requiring simultaneous segmentation of organs and tumors from medical images with a high class imbalance.
\subsubsection{Bladder tumor dataset} The bladder tumor dataset contains 2200 MRI slices from 25 patients with pathologically confirmed bladder cancer. A high-resolution Axial T2-weighted (T2W) MRI sequence was adopted. The imaging process contained from 80 to 124 slices per scan, each of size 512×512 pixels, with a pixel resolution of 0.5 × 0.5 $m m^{2}$. For each MRI scan, both bladder wall and tumor regions were manually delineated by an expert. Particularly, during the delineation process, all target regions were outlined slice-by-slice by the expert who was blinded to the pathological results of patients.
\subsubsection{Pancreas tumor dataset} The pancreas tumor dataset is a sub-dataset of the Medical Segmentation Decathlon (MSC) MICCAI 2018 challenge. It comprises 282 portal venous phase CT scans for training. An expert abdominal radiologist annotated the pancreatic parenchyma and pancreatic mass (cyst or tumor) in each slice. Please refer to~\cite{simpson2019large} for more details.

\subsection{Implementation Details}
\subsubsection{Data Pre-processing}
We first extract slices from the 3D scans along the axial plane. All 2D slices were normalized to $[0,1]$, and resized to $512\times512$ pixels. To prevent extra noise from the interpolation operation, we did not use any data augmentation operations.
\subsubsection{Training}
Our network was trained using Adam optimizer with an initial learning rate of 0.0001 and a batch size of 12. All datasets were randomly divided into 5 folds, with each fold been tested while the remaining data are further split into training set (75\%) and validation set (25\%). The experiments were performed on two NVIDIA GTX 1080 Ti GPU with a total of 22 GBs of graphics memory. One fold training takes about 12 hours for bladder dataset and 24 hours for pancreas dataset.
\subsubsection{Evaluation Metrics}
To evaluate segmentation performance, we adopted the common Dice Similarity Coefficient (DSC) and Jaccard Similarity Coefficient as the quantitative metrics.

\begin{table*}[]
\centering
\caption{Dice and Jaccard similarity coefficient (\%) of bladder wall and bladder tumors ($mean \pm standard\ deviation$).}
\label{table1}
\begin{tabular}{llllll}
\hline
\multirow{2}{*}{Method} & \multicolumn{2}{c}{Bladder Wall} &  & \multicolumn{2}{c}{Bladder Tumors} \\ \cline{2-3} \cline{5-6}
                        & Dice            & Jaccard        &  & Dice             & Jaccard         \\ \hline
UNet baseline~\cite{ronneberger2015u}         & $88.34\pm11.55$           & $80.56\pm13.84$         &  & $73.12\pm30.60$           & $64.62\pm29.50$          \\
Dilated UNet           & $89.05\pm10.34$           & $81.42\pm12.89$          &  & $75.40\pm27.72$            & $66.72\pm27.12$           \\
Auto-Focus~\cite{qin2018autofocus}             & $88.91\pm12.05$           & $81.32\pm14.55$          &  & $69.46\pm28.77$            & $60.76\pm27.95$           \\
Attention UNet~\cite{oktay2018attention}        & $88.74\pm9.53$           & $80.93\pm11.88$          &  & $73.76\pm30.47$            & $65.25\pm29.61$           \\
Ours                    & \bm{$89.32\pm10.19$}  & \bm{$81.82\pm12.59$} &  & \bm{$80.38\pm22.60$}   & \bm{$71.48\pm23.29$}  \\ \hline
\end{tabular}
\end{table*}

\begin{table*}[]
\centering
\caption{Dice similarity coefficient (\%) of normal pancreas tissue and pancreas tumors ($mean \pm standard\ deviation$).}
\label{table2}
\begin{tabular}{lccc}
\hline
Method  & Categorization & Pancreas Dice & Pancreas Tumors Dice \\ \hline
3D UNet & 3D             & $79.20\pm9.43$         & $35.61\pm32.20$          \\
VNet    & 3D             & $79.01\pm9.44$         & $35.99\pm31.27$                \\
V-NAS~\cite{Zhu2019VNASNA}   & Search         & \bm{$79.94\pm8.85$}        & $37.78\pm32.12$               \\
nnUNet\_2D~\cite{isensee2018nnu} & 2D             & 74.70         & 35.41                \\
nnUNet\_3D~\cite{isensee2018nnu} & 3D             & 77.69         & 42.69                \\
nnUNet\_3D Cascade~\cite{isensee2018nnu} & 3D Cascade     & 79.30         & \textbf{52.12}                \\
Ours    & \textbf{2D}             & $77.01\pm10.47$         & $50.12\pm30.86$                \\ \hline
\end{tabular}
\end{table*}

\begin{table}[]
\centering
\caption{Mean Dice similarity coefficient (\%) of bladder and pancreas. Label 1 (normal tissues) and 2 (tumors). Cd represent Coord}
\label{table3}
\begin{tabular}{l|cc|cc}
\hline
\multicolumn{1}{c|}{}                & \multicolumn{2}{c|}{Bladder} & \multicolumn{2}{c}{Pancreas} \\
label                                & 1             & 2            & 1             & 2            \\ \hline
Deform UNet (without local path) & 88.85         & 75.10      & 76.12         & 47.26       \\
Deform UNet (plain Conv)             & 89.44         & 74.30         & \textbf{78.01}& 42.84        \\
Deform UNet (Cd Conv)              & 89.23         & 74.98         & 77.24         & 45.62      \\
Deform UNet (Cd Pool)              & \textbf{89.57}& 76.93         & 76.58         & 48.87             \\
Deform UNet (Cd Conv/Pool)    & 89.32         & \textbf{80.38}& 77.01         & \textbf{50.12}        \\ \hline
\end{tabular}
\end{table}

\subsection{Results}
We compare our PGD-UNet with recent UNet-based improvement methods on bladder datasets, and report results on a 5-fold cross validation evaluation in Table~\ref{table1}. Our PGD-UNet achieves the best performance for both bladder and tumor segmentation. In particular, compared to the original UNet, PGD-UNet obtains a moderate improvement in bladder wall segmentation, whereas it achieves a significant improvement in bladder tumor segmentation. This indicates that our approach is robust to irregular shape variations, especially for tumors. Experiments of pancreas tumor segmentation are compared to the reported state-of-the-art methods on Medical Segmentation Decathlon (MSC) datasets in Table~\ref{table2}, where the `Categorization' column represents the type of method, `Search' refers to the method of automated network architecture search and 'Cascade' refers to the multi-stage method. Our PGD-UNet obtains comparable segmentation accuracy to the state-of-the-art 3D methods with a much simpler 2D network that requires less computational power and does not rely on exhaustive annotations for the full 3D image volumes. Compared with other 2D model, i.e. nnUNet\_2D\cite{isensee2018nnu}, our method improves dice performance by 3.09\% and 41.54\% for pancreas and pancreas tumors, respectively. All results are given by $mean \pm standard\ deviation$ for each sample.

\begin{figure*}[]
\centering
\includegraphics[scale=0.43]{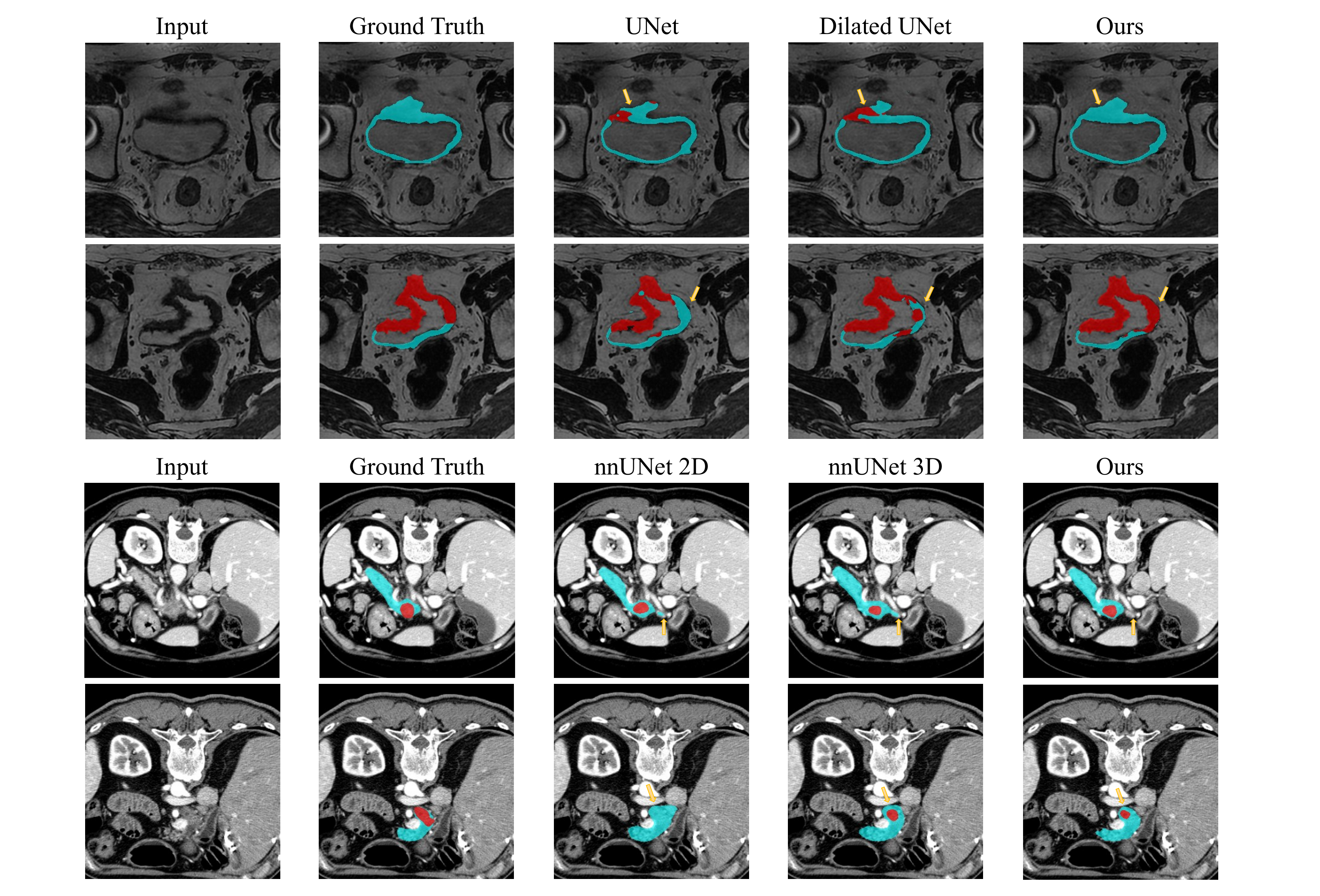}
\caption{Input, ground truth and segmentation results from comparison methods for Bladder (top) and Pancreas (bottom) datasets. Cyan indicates organ, red indicates tumor, and yellow arrows highlight the structures improved by our PGD-UNet}
\label{res}
\end{figure*}

We visualize some segmentation instances resulted from different algorithms on both datasets in Fig.~\ref{res}. As seen from the results, PGD-UNet is able to learn the discriminative features that can effectively segment narrow structures like bladder wall and complex pattern of tumors with varying shapes and sizes. Segmentation details in areas highlighted in organ also indicates that our method can effectively deal with boundary regions where tumors and bladder wall mix together.

\begin{table}[]
\centering
\caption{Ablation of loss function (mean DSC). Label 1 (normal tissues) and 2 (tumors).}
\label{table4}
\begin{tabular}{l|cc|cc}
\hline
\multicolumn{1}{c|}{}                & \multicolumn{2}{c|}{Bladder} & \multicolumn{2}{c}{Pancreas} \\
label                                                       & 1             & 2             & 1             & 2            \\ \hline
$\mathcal{L}_{\textit{FL}}$                                 & 89.54         & 77.05         & \textbf{78.95}& 45.48             \\
$\mathcal{L}_{\textit{GHM}}$                                & 86.37         & 73.23         & 72.77         & 25.67             \\
$\mathcal{L}_{\textit{DSC}}$                                & 81.15         & 48.29         & -             & -             \\
$\mathcal{L}_{\textit{FL}} + \mathcal{L}_{\textit{DSC}}$    & \textbf{89.97}& 75.59         & 75.95         & 48.81             \\
$\mathcal{L}_{\textit{NSFL}}$                               & 89.32         & \textbf{80.38}& 78.11         & 46.32             \\
$\mathcal{L}_{\textit{NSFL}} + \mathcal{L}_{\textit{DSC}}$  & 88.31         & 70.91         & 77.01         & \textbf{50.12}        \\ \hline
\end{tabular}
\end{table}

\subsection{Ablation Experiments}
The ablation experiments are performed to verify the contribution of each proposed module.
\subsubsection{Localization Path}
We compared the performance of the model with and without localization path, and carried out ablation experiments on important components of `CoordConv' and `CoordPool'. As shown in Table~\ref{table3}, segmentation performance degrades significantly when removing the localization path. The second row represents a localization path consisting of plain convolutions. Comparing the second and following rows, it can be seen that using CoordConv alone has only a slight effect, whereas the CoordPool that preserves position information impacts more on the DSC. In addition, the results in the last row show that localization path improves the segmentation accuracy of tumor much more than that of normal tissues. This is consistent with the observation that tumors have more size and shape variations than normal tissues.

\begin{figure*}[]
\centering
\includegraphics[scale=0.4]{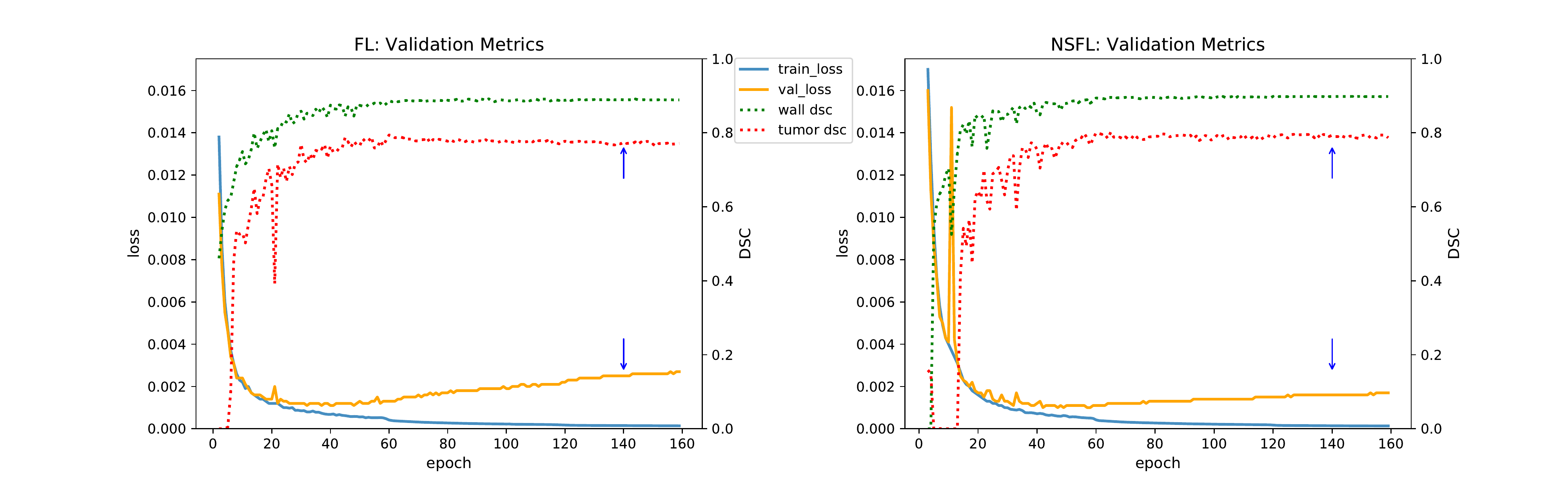}
\caption{Loss value and DSC curve for focal loss and noise suppression focal loss on MRI bladder dataset. Blue arrows point to the boundaries of loss and tumor dice of validation at epoch 140.}
\label{curve}
\end{figure*}

\subsubsection{Noise Suppression Focal Loss}
Due to the large proportion of background in our datasets, using the Cross-Entropy (CE) loss function alone cannot make network converge, and all the outputs predict the background as results. In this case, we chose Focal Loss (FL) as the baseline. Besides, other loss functions that aiming at handling class imbalance were compared, including Gradient Harmonizing Mechanism (GHM) loss, DSC loss and their combination.

Table~\ref{table4} reports the results of ablation experiments using various loss function on the bladder and pancreas datasets. The DSC of tumor consistently increases by adding the NSFL, whereas the performance of normal tissue degrades slightly. This indicates that the impact of NSFL positively relates to the level of label noise. Using the DSC loss alone is unstable and may cause a sharp decline in tumor segmentation performance. We believe that this is due to the class imbalance between normal tissue and tumor. As DSC loss is based on regional integration, the classes with abundant pixels are prone to dominate the gradient, thus leading to poor results for other classes or even failing to converge.

Fig.~\ref{curve} compares the evolution of loss value and validation metrics between FL and NSFL on MRI bladder dataset. After 50 epoch, the validation set loss of FL began to rise, indicating the overfitting of the network. Meanwhile, NSFL suppressed this trend significantly. Besides, as can be seen from the curve of DSC metrics on the validation set, normal tissues hardly to overfit due to the large number of samples and clean label, whereas tumors are prone to overfit. Thus, NSFL helps to reach the optimal convergence point for both normal tissues and tumors achieving precise segmentation results.

\section{Conclusions and Future Work}
We proposed an improved UNet framework named PGD-UNet for medical image segmentation. PGD-UNet enhances the original UNet by including deformable convolution with localization path and noise suppression focal loss function to effectively address the problem of size and shape variations, and severe class imbalance in tumor segmentation. By adding `CoordConv' and `CoordPool' modules, we explicitly encode position information into the network to improve the offset learning of deformable convolution. To solve the problem of confusion between noise and hard-to-classify samples caused by focal loss when applying it to deal with class imbalance, we design a new loss function to suppress the impact of outliers on the gradient. The effectiveness of our method is verified on two challenging medical segmentation tasks. In the future, we plan to extend our work to allow utilising complementary information from both MRI and CT images, where challenges associated like registration \cite{gong2016} need to be solved.

\section*{Acknowledgment}
This work was supported in part by the National Natural Science Foundation of China (NSFC) under Grant No. 61671151 and 61573097, the Natural Science Foundation of JiangSu Province under Grant No. BK20181265, the Australian Research Council (ARC) under Grant No. LP170100416, LP180100114 and DP200102611, and the Research Grants Council of the Hong Kong SAR under Project CityU11202418.

\bibliographystyle{IEEEtran}
\bibliography{IEEEabrv,conference_101719}

\end{document}